# More Jabber about the Collatz Conjecture and a Closed form for Detecting Cycles on Special Subsequences.

Thomas W. Lynch
2011 06 02

## Introduction

In [ref 1] apparently published in 2006 professor Cadogan presents an interesting function that maps natural numbers into a two dimensional array based on how many ones the number trails off with. The length of trailing one's is said to be the row index. The rest of the number, divided by 2, is the column index. He then shows that applying the Collatz function reduces the ending string of ones by one causing it to drop down a row. At the bottom of the table are odd numbers ending in 'xxx..01' - a tail length of one. He reports that a restart function and continued iteration then causes a to monotonic reduction in column indexes, and the existence of this algorithm proves the conjecture.

This paper provides a bit more formal framework as a basis for further discussion. Herein I closely examine the functions in involved, provide some nomenclature, and examine the general concept of column reduction. In contrast to prior results, this paper purports that no algorithm which makes use of only row index reduction termination points can display the trait of monotonic column reduction.

However, the concept of a row reduction does more than just lead to a column termination point, it also identifies strands in the more general Collatz sequences that have a special property, that of the existence of a unique inverse function. This makes analysis on these strands simpler than for the sequence in general. By using this property I derive a closed form formula which when evaluated would identify cycles in such strands. However, it involves factorization and mixed integer solutions of exponential functions. Cycles on the strands, if they exist, appear to only exist for very large numbers.

## 1 Collatz's Conjecture

Unless otherwise stated all variables used in this paper are non-negative integers i.e. Natural numbers. I use $\boldsymbol{N}$ for the set of natural numbers, $\boldsymbol{O}$ for the set of odd numbers, $\boldsymbol{E}$ for events, and $\boldsymbol{Q}$ for rationals. A superscript on a function name represents repeated applications of the function. I don't use powers of functions in this paper.

Let us start with the Collatz function:

*def 1.1; Collatz Function*: $\gamma(n)=(3n+1)2^{-k} \wedge pick \;\; k \mid \gamma(n) \in \boldsymbol{O}$

To calculate the result of this function from the given input $n$ we multiply $n$ by 3, add 1, and then divide the result repeatedly by 2 until the result is an odd number. The number of times we divide by two is in fact another calculated result, denoted as *k*.

def 1.1.1 We say $k = \|\gamma(n)\|$

It follows from the definition that $\gamma(n)$ can not produce an even number. In addition it can not produce a multiple of 3. This can be shown by the fact there are exceptions to possible inputs to the inverse function, see *lemma 1.1* below.

We may now simply state the Collatz conjecture:

*def 1.2; Collatz Conjecture*: $\exists_m : \gamma^m(n) = 1$

The Collatz conjecture supposes that there always exists a finite number of applications of the Collatz function that will reduce any given positive integer to 1.

As an example, the Collatz conjecture holds for the value 0 with m=1. As *m* is one, the Collatz function is applied only once, and the parameter *k* for that application will be zero.

The inverse of the Collatz function maps odd numbers back to natural numbers. This function has one explicit input, say $n \in \boldsymbol{O}$ , and an implicit input *k*. Because the result is a natural number, *k* must make $n2^k - 1$ divisible by 3. Hence,

*lemma 1.1; Collatz inverse*: $\gamma^{-1}(k, n) = (n2^k - 1)3^{-1}$

But suppose we had an odd number, say *x*, for which no *k* existed that would produce a result $\gamma^{-1} \in \boldsymbol{N}$ , that would imply $\neg\exists_k x2^k - 1 = 0 \mod 3$

We can break the proof down into three test cases as $x \bmod 3$ is either 0,1, or 2:

| | | |
|---|---|---|
| case $x = 0 \bmod 3$ | $x2^k - 1 = 2 \mod 3$ | as x mod 3 is zero, we are left with -1 |
| case $x = 1 \bmod 3$ | $x2^k - 1 = 2^k - 1 \mod 3$ | holds for k even, as $2^{2n} = 1 \mod 3$ |
| case $x = 2 \bmod 3$ | $x2^k - 1 = 2^{k+1} - 1 \mod 3$ | holds for k odd, as $2^{2n} = 1 \mod 3$ |

We discover that for the case of $x = 0 \bmod 3$ *k* drops out and we are left with a constant of 2. No value of *k* can cause this to be a multiple of 3, so :

*lemma 1.2;* $\gamma(n)$ can not produce numbers that are multiples of 3.

## 2 The Exception Set

Lets consider a set of non-negative integers that are exceptions to the Collatz conjecture.

Set of exceptions to the Collatz Conjecture: $E_0 = \{n \mid n \in \boldsymbol{N} \wedge \neg\exists_m : \gamma^m(n) = 1\}$

This says that for each exception, *n*, no matter how many times we apply the Collatz function to *n*, we will never converge to the number 1. This set will be $\emptyset$ if and only if the Collatz conjecture is true.

We know we don't need to test the number 0 for inclusion in this set because the Conjecture holds in that case. Also we don't need to include the number 1, as that is the definition of convergence. So we can make our definition more expressive without danger of having removed an exception from the set :

*def 2.1;* $E_0 = \{ n \mid n \in \mathbf{N} \wedge n > 1 \wedge \neg \exists_m : \gamma^m(n) = 1 \}$

lemma 2.1; No Need to Test Even Numbers as Exceptions

By definition the result of applying the Collatz function is odd. Thus if there is an even number, $n$ ,that is an exception, then the odd number $\gamma(n)$ must also be an exception. Consequently the set will not become empty if we take out the even numbers.

$E_1 = \{ n \mid n \in \mathbf{O} \wedge n > 1 \wedge \neg \exists_m : \gamma^m(n) = 1 \}$

We may use an analogous argument to show we do not need to test multiples of three.

$E_2 = \{ n \mid n \in \mathbf{O} \wedge n \bmod 3 \neq 0 \wedge n > 1 \wedge \neg \exists_m : \gamma^m(n) = 1 \}$

# 3. Professor Cadogan's Function

[ref 1] defines a number as a function of two parameters:

*def 3.1;* $\forall_{i \geq 1, j} : c^{-1}_{i,j} = j 2^{i+1} + 2^i - 1$ in [ref 1]: $t_{i,j} = j 2^{i+1} + 2^i - 1$

*lemma 3.1* $c(n) = (i, j)$

c(n) maps the set of odd numbers, ***O***, to $\mathbf{N} \times \mathbf{O}$ : $\mathbf{O} \xrightarrow{c} \mathbf{N} \times \mathbf{O}$ .

We only consider odd valued $c^{-1}_{i,j}$ this implies that *i* will always be 1 or greater.

The next section shows that $c(n)$ exists and can be uniquely constructed for any odd natural number.

### *Constructing c(n) from Any Odd Natural Number*

The parameter *i* will be determined while evaluating the *tail* function. The parameter *j* will be determined while evaluating the *head* function as described in this section.

### Construction of the *tail(n)* and determination of *i*

Given an odd integer, *n,* consider that number to be represented in binary notation. Now look at the least significant (right most) bit of the number. It is a one. This has to be the case as we are looking at an odd number. Consider this 1 to be the right most bit of the *tail*. Then scan left bit by bit until a zero is discovered. All bits to the right of the zero are part of the *tail*. The parameter *i* is the number of ones placed in the tail. As we have an odd number the smallest value for *i* will be one. It follows that the tail will have a numerical value of $2^i-1$ .

*def 3.3:* for any odd *n* pick the largest *i* such that $n-(2^i-1)=2^i\lfloor n2^{-i}\rfloor$ then $2^i-1$ is the $tail(n)$ and *i* is said to be the length of the tail, also $i=\|tail(n)\|$

Again the subscript on the function name is used to indicate an implicit result, here it is called *i*. Often the subscript will be dropped.

*lemma 3.3;* $tail(n)$ will always be an odd natural number less than or equal to *n*.

The operations used to create *n* are closed on the integer field for non-negative *i*. *n* itself is a positive odd natural number. As such, $\lfloor n2^{-i}\rfloor$ will be positive or zero. As *n* is an odd natural number, at least 1 can always be subtracted from it without the result becoming negative, so *i* is an integer greater than or equal to one. As powers of two are even, $2^i-1$ is odd. It is a consequence of the definition that subtracting *tail* from *n* will leave a zero or positive value, therefore *tail* is less than or equal to *n*.

### Construction of the *head(n)* and determination of *j*

In the definition of the tail we referred to the floor $\lfloor n2^{-i}\rfloor$ this number will be even. This is a consequence of the fact that the *tail* function stripped off all the bottom ones. If the left over after this operation were odd, there would be more to ones to strip off. Hence we define the head as:

*def 3.4* $head(n)=\lfloor n2^{-(i+1)}\rfloor$

This value could be any natural number, including zero. It follows that:

*lemma 3.4* $n=head(n)2^{i+1}+tail(n)=c^{-1}(c(n))$ substitute *def 3.3* and *3.4* into *def 3.1*

*def 3.5* when $c(n)\rightarrow(i,j)$ we say that $j=head(n)$

*lemma 3.5;* This section has demonstrated a constructive procedure for converting any odd natural number into *(i,j)* , so *c(n)* is defined for all odd natural numbers.

## *A Table of n Placed at Coordinates c(n)*

For each number, *n*, in this table, *i* is the row index and *j* is the column index (from [ref 1]) . Note that the value 1 appears at location (1,0).

| | j=0 | 1 | 2 | 3 | 4 | 5 |
|---|---|---|---|---|---|---|
| i=1 | 1 | 5 | 9 | 13 | 17 | 21 |
| 2 | 3 | 11 | 19 | 27 | 35 | 43 |
| 3 | 7 | 23 | 39 | 55 | 71 | 87 |
| 4 | 15 | 47 | 79 | 111 | 143 | 175 |
| 5 | 31 | 95 | 159 | 223 | 287 | 351 |

*table 3.1*

### ***The value 1 uniquely appears at location (1,0)***

$$c^{-1}(i, j) = j2^{i+1} + 2^i - 1 \qquad \text{def 3.1}$$

As already noted, the minimum *i* for an odd *n* is 1. The parameter *i* only occurs in positive terms of the equation shown in *def 3.1*, indeed only as a power of 2. Hence, $c^{-1}(i, j)$ increases strictly monotonically with increasing *i*. Likewise *j* only appears in a positive term as a coefficient of a non-zero value. Hence $c^{-1}(i, j)$ increases strictly monotonically with *j*. When *i* is 2, and *j* is 0, $c^{-1}(i, j)$ is already larger than 1. When *j* is 1 and *i* is at its minimum value of 1, $c^{-1}(i, j)$ is already larger than 1. It follows that $c^{-1}(i, j)$ *= 1* can only occur when *j=0* and *i=1.*

*lemma 3.6;* The number 1 only occurs at location (1,0) in the table.

### **Relationships between *tail* and *k***

Multiplying a string of ones by three is the same as left shifting one number by 1, and then adding it back into itself. However, we can take the number to be added and use it to fill in the zero caused by the left shift. Hence, we will end up with a string of ones added to a string of ones. For the bottom of the number this is the same as multiplying by 2. It follows that k will have to be *1* in order to normalize the result. Though this section appears towards the front it was added last, so I hope you will forgive the ASCII graphics, here are some examples:

(2 + 1) tail + 1 = (2 tail + 1) + tail

```
011110
h01111
---------
b01101
+       1
---------
b01110 --> k=1
```

```
h011
011
------
b011
+    1
------
b110 --> k=1
```

hh01
h01
------
0h11
+   1
?b00

---> woops, for a tail length of 1, *k* is related to head

*lemma 3.7;* For numbers in rows greater than 1, $k=1$ .

$$\forall_{i>1}:\ \|\gamma(c^{-1}(i,j))\|=1$$

Corollary 3.7.1

$$n=c^{-1}(i,j) \wedge \|tail(n)\|>1\ \rightarrow\ \|\gamma(n)\|=1$$

Corollary 3.7.2

$$c(n)_i>1\rightarrow\|\gamma(n)\|=1$$

Here the subscript *i* on *c(n)* means we want the *i* component of the vector result, *(i,j).*

*lemma 3.7.3;* Due to lemma 3.7 we can define a one operand inverse Collatz function

The value for *k* that would take one back to a prior value, say $n_0$ , from a current value, say $n_1$ , is identical to the value *k* that calculate by $\gamma$ when going from from $n_0$ to $n_1$ in the first place. This is because the inverse was derived algebraically for lemma 1.1, so the k in both functions is in fact the same number.

$k=\|\gamma(n0)\|$ ; by *def 1.1.1*

$$n_0=\gamma^{-1}(\|\gamma(n_0)\|,n_1)$$

Hence, when the inverse is to go into the table, other than to row 1:

$$\gamma^{-1}(n)=(2n-1)3^{-1}$$

Inverse paths are unique for *c(n)* with row index greater than 1. There are the other inverse paths, indeed there are k of them and k is unbounded. These other paths must be in row 1.

This inverse is still only defined over natural numbers when $n \bmod 3=2$ It is defined for row 1 numbers, but will take them back up to higher row numbers rather than following any path back into row 1.

## 4 Row Index Reduction in $c(n)$ against $\gamma$

*lemma 4.1;* Applying the Collatz function, $\gamma$ , to a number on row $i_0$ for all $i_0>1$ , results in a number on row $i_0-1$ [ref 1].

We can derive this result algebraically in a similar manner as done in [ref 1] while using our *head* and *tail* definitions as a device:

| | | |
|---|---|---|
| | $\gamma(head(n)2^{i0+1}+tail(n))$ | inserting eq 3.5 |
| *term 4.1* | $(3(head(n)2^{i0+1})+3\,tail(n)+1)2^{-k}$ *; chose k so result is an odd integer* | definition $\gamma$ |

Now consider what happens when the tail is multiplied by 3 and 1 is added:

| | | |
|---|---|---|
| | $3\,tail(n)+1$ | taken from *term 4.1* |
| | $2\,tail(n)+tail(n)+1$ | 3 = (2 + 1) |
| | $2(2^{i0}-1)+(2^{i0}-1)+1$ | apply def *tail* |
| | $2^{i0+1}-2+2^{i0}$ | collect terms |
| | $2^{i0+1}+2(2^{i0-1}-1)$ | separate out form of a tail |
| *term 4.2* | $2^{i0+1}+2(2^{i1}-1)$ | say that $i_1=i_0-1$ |

Substituting this back into *term 1*:

| | | |
|---|---|---|
| | $(3\,head(n)2^{i0+1}+3(2^{i1}-1)+1)2^{-k}$ | *term 4.1* again |
| | $(3\,head(n)2^{i0+1}+[2^{i0+1}+2(2^{i1}-1)])2^{-k}$ | substitute in *term 4.2* |
| | $([3\,head(n)2^{i0+1}+2^{i0+1}]+2(2^{i1}-1))2^{-k}$ | regrouping |
| | $([3\,head(n)2^{i0}+2^{i0}]+(2^{i1}-1))2^{-k+1}$ | adjust the scale by one |
| | $([3\,head(n)2^{i1+1}+2^{i1+1}]+(2^{i1}-1))2^{-k-1}$ | apply def $i_1=i_0-1$ |
| | $([3\,head(n)+1]2^{i1+1}+(2^{i1}-1))2^{-k-1}$ | pull $2^{i1+1}$ out of [] |
| *term 4.3* | $head(\gamma(n))2^{i1+1}+tail(\gamma(n))2^{-k-1}$ | $i_1=i_0-1$<br>$head(\gamma(n))=3\,head(n)+1$<br>$tail(\gamma(n))=2^{i1}-1$ |

*lemma 4.2;* Recall that *j* is identical to *head(n)*. So $head(\gamma(n))=3\,head(n)+1$ means the same thing as $j_1=3\,j_0+1$ .

Which leads to:

$$c(n)=(i,j)\rightarrow c(\gamma(n))=(i-1,3j+1)$$

| | | |
|---|---|---|
| *lemma 4.3;* | $(i,j)\overset{\gamma}{\rightarrow}(i-1,3j+1)$ | restatement of prior line, from [ref 1] |

The nifty thing about this result is that it shows we can calculate changes in *i* and *j* against applications of $\gamma$ without having to explicitly invoke Cadogan's function.

### *Behavior of Row Index i under Repeated Applications of* $\gamma$

*lemma 4.4;* It follows from lemma 4.1 that:

$$\forall_{(i_0 \geq 1)\wedge(m \leq i_0 - 1)} : i_0 \xrightarrow{\gamma^m} i_0 - m$$

Hence there always exists a Collatz Sequence leading to $c(n)=(1,j)$ . [ref 1]. As for the first term in the quantification, by definition all row indexes, i, are greater than or equal to one. Since a row index can not be less than one, *m* cannot be greater than $i_0 - 1$ ;

### *Behavior of Column Index j under Repeated Applications of* $\gamma$

*Lemma 4.1* establishes an increase in parameter *j* against the application of the Collatz function. We will call this the function *g*.

def 4.1 $g(j)=3j+1$ term from lemma 4.2

I find it interesting that the column expansion function $g$ has nearly the same form as the Collatz function $\gamma$ . The only difference is that $\gamma$ normalizes with a right shift, $2^{-k}$ , where as $g$ does not.

lemma 4.5 $g(j) \geq \gamma(j)$

Note here in lemma 4.5, we are doing an unusual thing and applying gamma to a column in index rather than a number from the table.

If the number had been from the table, other than in row 1. we would know k to be 1 due to *lemma 3.7* In which case we would have:

corollary 4.5.1 $i>1 \wedge n=c^{-1}(i,j) \rightarrow g(n)=2\gamma(n)$

This time we have done the unusual thing of applying *g* to a number from the table, as usually it is applied to a column index, but a function doesn't care where its operands came from.

By algebraically manipulating the definition of $g(j)$ to form an inverse we get:

lemma 4.6 $g^{-1}(j)=(j-1)3^{-1}$ derived algebraically from def 4.1

Over the domain of natural numbers *g* only has an inverse for certain values of *j* which are 1 larger than a multiple of 3.

*def 4.2;* we extend the definitions of $g^{-1}$ and $g$ to operate over the rational field.

*g* is closed over natural numbers, but it may or may not produce a natural number given a rational input. We can say this:

*lemma 4.7;* $\forall_{j,m,n \in \boldsymbol{N} \wedge m \geq n} : g^n(g^{-m}(j)) \in \boldsymbol{N}$

When *n* is equal to *m* we get back *j*. There may have been some rational intermediate values, but those get undone again. When we continue after getting back to the starting point by making *n* larger then *m* we are working with *j*, a natural number, and *g* is closed over natural numbers, so the results from all further applications of *g* will be in $\boldsymbol{N}$ .

### ***The Behavior of*** $k$ ***relative to*** $g^{-1}$

In lemma 4.3 the result of applying $\gamma$ to a number $n_0$, located at the point $c(n_0)$, caused the $j_0$ component of $c(n_0)$ to go to $3j_0+1=j_1$. We called this column index transformation function g. In other words:

$$g(c(n_0)_j)=c(\gamma(n_0))_j$$

Note the subscript *j* on $c(n)_j$ means we want the *j* component from the vector result *(i,j)*. Then we derived an inverse function that would take us back again. $g^{-1}(j_1)=j_0$ The interesting part about this inverse is that it accepts only one parameter while $\gamma^{-1}$ requires two parameters.

$$g^{-1}(c(n_1)_j)=c(\gamma^{-1}(k,c(n_1)_j))_j$$

This implies that built into the definition of $g^{-1}$ is an assumption about the value *k* that would be put into $\gamma^{-1}$ to cause it to follow one specific backwards path. This value of *k* must be the same k that was used for normalization in the forward direction, namely $\|\gamma(n_0)\|$, and indeed, due to lemma 3.7, we know *k* to be 1 $\forall_{i>1}\|\gamma(c^{-1}(i,j))\|=1$

### ***Iteration***

Now lets consider what repeated applications of *g* does to a $j_0$ If we write the sequence of $j_m$ resulting from repeated applications, in base 3 number representation we we get the pattern:

| in base 3 | in base 10 | as g application | application count |
|---|---|---|---|
| $j_0 1_3$ | $3j_0+1$ | $g(j_0)$ | m=1 |
| $j_0 11_3$ | $9j_0+4$ | $g^2(j_0)$ | m=2 |
| $j_0 111_3$ | $27j_0+13$ | $g^3(j_0)$ | m=3 |

We get a head followed by a tail of 1s. In this case the tail is of ones in base 3. In any base subtracting 1 from the $b^n$, will produce a string of digits of value $b-1$. In this case it will be a sequence of 2s. Of course a sequence of 2s divided by two is a string of ones. Hence we get:

Summation form:

$$j_m=j_0 3^m+\sum_{n=0}^{m-1}(3^n)$$ creating the string of 1s

$$j_m=j_0 3^m+\frac{1}{2}(3^m-1)$$

*lemma 4.8;* $$g^m(j)=3^m j+\sum_{n=0}^{m-1}(3^n)$$

*lemma 4.9;* $$g(j)^m=j3^m+\frac{1}{2}(3^m-1)$$

I find it interesting that $g^m$ has a head and a tail and is of similar form as $c^{-1}$. Perhaps this implies that $c^{-1}$ could also be derived as the repeatedly application of a simpler function.

*lemma 4.1;* combined with either lemma 4.8 or lemma 4.9 produces:

*lemma 4.10;* $(i,j) \overset{\gamma^m}{\rightarrow} (i-m, 3^m j + \sum_{n=0}^{m-1} (3^n))$

*lemma 4.11;* $(i,j) \overset{\gamma^m}{\rightarrow} (i-m, 3^m j + \frac{1}{2}(3^m - 1))$

### About Row 1 Numbers

This result is presented in [ref 1]. By definition in row 1, $i=1$. Hence the tail is one bit long, scanning to the left this bit is followed by a zero, and then there is the head of the number, *j*.

*lemma 4.12;* $c^{-1}(1,j) = 4j+1$ *def 3.1* evaluated with i = 1

Hence, all numbers in the first row have values that can be calculated simply from their column index - which could in fact be said of numbers on any row. Rather we discuss row 1 because of this surprising property:

*lemma 4.14;* $\gamma(4j+1) = \gamma(j)$

This says that applying the Collatz function to a row 1 number yields the same result as applying the Collatz function to the column index. The derivation goes like this:

$$\gamma(4\text{n}+1) = (3(4n+1)+1)2^{-k} = (12n+3+1)2^{-k} = (3n+1)2^{-k+2}$$

$$\gamma(n) = (3n+1)2^{-k}$$ same form as above

Here are a couple of examples:

$$5 \overset{\gamma}{\rightarrow} (5 \cdot 3+1) 2^{-4} = 1$$

j=5
4 j+ 1 = 21

$$21 \overset{\gamma}{\rightarrow} (21 \cdot 3+1) 2^{-6} = 1$$

$$2 \overset{\gamma}{\rightarrow} (2 \cdot 3+1) 2^{0} = 7$$

j=2;
4 j + 1 = 9

$$9 \overset{\gamma}{\rightarrow} (9 \cdot 3+1) 2^{-2} = 7$$

This is a strange result, as it says we can exchange a number in row one of the table for its column index and continue iterating. While numbers in the table are of special form, the column index can be anything, including an even number.

If the column index has a $tail(j)>1$ , then it has a corresponding entry in the table at a row index greater than one, and thus has the form $c^{-1}(\|tail(j)\|, head(j))$ . In such a case $head(j)$ , the new column number, will be smaller than $j$ by at least a factor of $2^{-i}$ , due to removing the tail to get to the head. However, recall that there was growth of $g^m$ to get to the termination column in the first place, which is of order $3^m$ . Which term dominates is a question to be answered further below in this paper.

If the column index for the termination point, j, has a tail of length 1, then once again it is a number of the form $4j+1$ , so we again land back on row 1. Though this time with a smaller j by a factor of 1/4.

If the column index for the termination point j is even, then we move to this even numbers odd partner after one application, which again could be any number. The zeros on the bottom of such a number multiplied by 3 during the Collatz function, remain zero. Adding 1 then makes the number odd. So the number increased in size by a factor of 3.

### *Iteration form for Row Index Reduction and Summary*

To recap, we start with a number, say $n_0$ , we apply the Cadogan function $c(n_0)$ to get $i_0$ and $j_0$ . We apply the Collatz function repeatedly, $\gamma^m$ until $i_m=1$ . We are guaranteed a monotonic incremental arrival in row index reduction due to *lemma 4.1*.

Lets give row reduction a name:

lemma 4.8 $\rho((i_{0,} j_0)) = g^{i_0 - 1}(j_0)$ which can also be written $\rho(c(n_0)) = g^{i_0 - 1}(j_0)$

$\rho$ accepts a start point and produces the termination column. By definition the termination row is the constant 1, so it need not be returned.

Due to lemma 4.3 we know we can calculate the $c(n_m)$ coordinates at each step $m \geq 1$ directly from the coordinates from the prior step at $c(n_{m-1})$. There is no need to evaluate $c(n_m)$ to find the coordinates. Indeed, due to *lemma 4.10* we can skip iteration altogether and just calculate $g^{i_0-1}(j_0)$ directly. We can also see that the column number *j* will be increasing as a power of 3 agatinst the number of steps from *lemma 4.9.* At each application of $\gamma$ the row index goes down by 1, so we know that the number of steps, *m*, needed to reach the termination point will be $i_0-1$.

Upon arriving at the termination point after row reduction we know from *lemma 4.12* that we are at the number, $n_{i_0-1}=4j_{i_0-1}+1$ Furthermore, we know from *lemma 4.14* that continuing from $n_{i_0-1}$ by applying $\gamma$ again, is the same as continuing from the column index itself $j_{i_0}$, i.e. that $\gamma(n_{i_0-1})=\gamma(j_{i_0})$.

Another thing we get from row reduction is an equivalence class of numbers, namely those that belonged to the 'row index reduction trajectory'. These numbers, apart from the last one in the trajectory, have long tails and related *head* values. Perhaps this causes other relational properties between them.

# 5 Repeated Row Reductions

Unless $\rho(c(n))$ happens to equal 0, the column for the number 1, the row reduction has left us with another number to apply the Collatz function to.

Now suppose we iterate row reductions, after each row reduction step *r*, we arrive at column, $j_{m,r}$. We will call this quantity more simply $J_r$.

*def 5.1*; $J_r=j_{m,r}$ where m will be $i_{0,r}-1$.

Successive row reductions will produce a sequence of the form $J_{0,}J_{1,}J_2..J_r$.

Row reduction transforms the question of the Collatz conjecture to the question of whether there exists a $J_r$ that is 0. (recall the number 1 is in row 1, column 0 by *lemma 3.6*) Hence,

*lemma 5.1;*The Collatz Conjecture holds if, and only if, $\exists_r : J_r \in Sequence(J_r) \wedge J_r=0$

Due to lemma 4.7 we have the option of getting the new start point by applying $\gamma$ to the number $J$ rather to the number $c^{-1}(1,J)$. Thus to start the next iteration we will find the length of $tail(\gamma(J))$ to set $i_0$ and the $head(\gamma(J))$ to set $j_0$.

*lemma 5.2*; $q(J)=(i_{0,}j_0)=(\|tail(\gamma(J))\|, head(\gamma(J)))$ ; calculating a start point for row reduction

corollary 5.1.1 Collatz holds if and only if:

$$\rho(J)=0 \ \vee \ [n_0=\rho(J) \wedge \exists_r (\rho \circ q)^r(n_0)=0]$$

### *The Convergence Boundary*

The numbers placed in the cells in table 1 are not necessary as they are implied by the coordinates, so we can use a dot diagram to describe c(n). This is good because *g* is exponential against step count, *m*, so we are going to need a lot of dots to create a useful visualization.

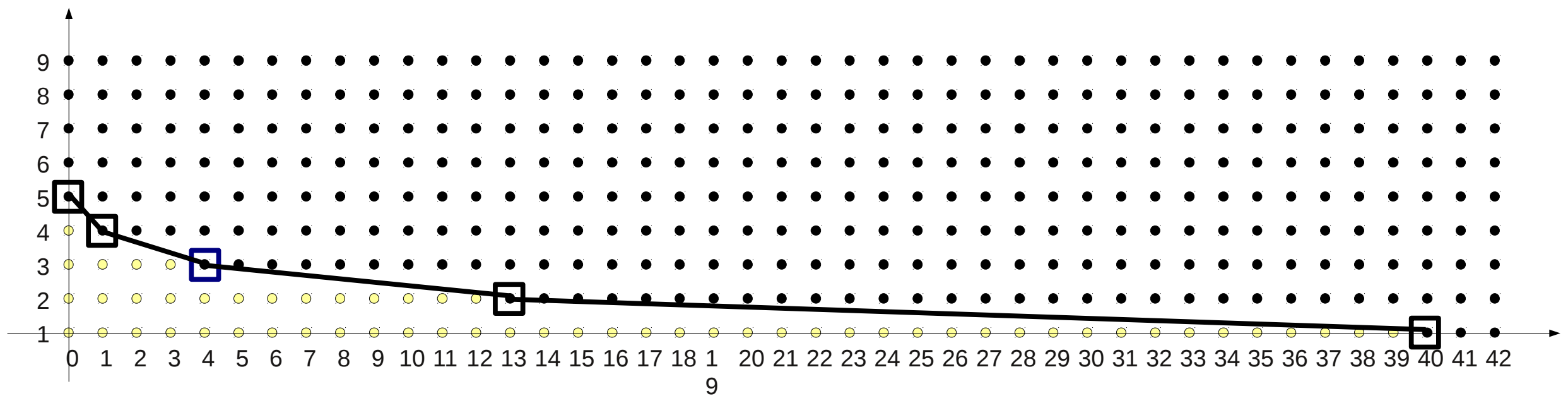

table 5.1

*Table 5.1* is a dot diagram view of *table 3.1*. It has been flipped vertically so that moving to lesser row indexes moves down the diagram. It has been extended so more cases can be shown. Here *i* stems from 1 to 9, and *j* stems from 0 to 42. Now the table looks more like a graph. Be careful to notice that coordinates are of the form (row, column) which are $(y, x)$ relative to the graph.

In table 2 we show a row index reduction trajectory starting from the number $n_0 = 111$, i.e. one hundred and eleven. $c(111) = (i, j) = (3{,}4)$. This trajectory is then shown going to the right, first arriving at $c(\gamma(111)) = (2{,}13)$ and then at the termination point $c(\gamma^2(111)) = (1{,}40)$.

We then go back to the start point (3,4) and use $g^{-1}$ to *complete* the trajectory by calculating trajectory crossing points on each successive row with increasingly higher indexes than our start point. This continues until we arrive at column 0. We first move up one row and find point (4,1), and then at the next step arrive at column 0 with (5,0). For this trajectory we say the *start point* for row reduction was (3,4) but the *first point* of the completed trajectory is (5,0). First points always have a column coordinate of zero. The termination point for both row reduction and for the trajectory is (1,40). Termination points always have a row coordinate of 1.

We happened to pick a case where $g^{-1}$ remained in the integer domain. However, for some start points when working backwards to complete the trajectory the column value for a row may be a rational value. In this case we treat our dot diagram as though it were laid over a $\boldsymbol{Q} \times \boldsymbol{Q}$ plane. The dots just mark special points on the plane, namely those with integral coordinates. Each row is interpreted as a line going horizontally. Similarly columns are lines going vertically. The trajectory has an intersection point with each row that occurs below or at the first point, though this intersection might not coincide with a dot. Similarly, the intersection where the trajectory crosses the first column (y-axis) might not have an integral row coordinate.

We will make use the term 'dot' to refer to a point that has integral coordinates.

lemma 5.3; definition of *lesser than, on,* and *greater than* the boundary

Completed row reduction trajectories will always extend from column 0 at the y-axis and then cross the plane while monotonically decreasing in value until crossing the x-axis at row = 1. The form of the curve is due to *lemma 4.11*. Hence, a completed trajectory bi-partitions the plane. For this reason we call it a *row index reduction boundary*, or just *boundary*. Dots on a given row that occur to the left of the boundary crossing are said to be *lesser than* the boundary, those to the right are *greater than*. Dots coincident to the boundary intersection are *on* the boundary. It may be the case for a given row that no dot lies on the boundary. All dots on rows above the *first point* of the boundary are said to be greater than the boundary. Due to shape of the boundary curve the first point will always be the highest point on a boundary.

*lemma 5.4*

Suppose we are given a boundary defined by completing a row reduction trajectory. Suppose we are also given an integral coordinate start 'dot' for another trajectory, say (i,j). a) If (i,j) is greater than the boundary all dots on the trajectory stemming from it remain greater than the boundary. b) If (i,j) is on the boundary, then all dots on the trajectory steaming from (i,j) are on the boundary. c) If (i,j) is lesser than the boundary then all points on the trajectory will be lesser than the boundary. In other words, all trajectories remain on the same part of the plane as their start points.

The proof of this is rather simple. Recall lemma 4.11,

$$(i,j)\overset{\gamma^m}{\rightarrow}\left(i-m,3^m j+\frac{1}{2}(3^m-1)\right) \quad \textit{lemma 4.11}$$

The column component calculation is independent of the row calculation as it does not include *i*. The row calculation just shifts vertically by a constant for different start points. Hence all completed row reduction trajectories are parallel. Of course, parallel trajectories do not cross.

*lemma 5.5:*

Trajectories stemming from lesser than dots have lesser than termination points.

This also follows from the fact that the trajectories are parallel and that the function is monotonically decreasing while starting from above the x-axis. When a trajectory is moved down, it intersects the x-axis sooner

*lemma 5.6 monotonic convergence;*

It follows from lemma 5.5, that a sufficient criteria to show the Collatz Conjecture holds is that the new start point for each next row reduction iteration always lies on the lesser side of the boundary set by the prior row reduction trajectory.

$$\forall_{J>0}:\ (q(J)=(i_{0,}j_0))<(i_{0,}g^{-(i_0-1)}(J))$$

Note that the term $g^{-(i_0-1)}(J)$ is the boundary intersection with row $i_0$, it is a column coordinate. At each application of $g^{-1}$ we walk up to the intersection on the next row, with one application taking us to the intersection with row 2, two applications taking us to row 3, etc. After $i_0-1$ applications we have the column intersection for row $i_0$. Due to the behavior of $g^{-1}$ (see the discussion of *lemma 4.6*) the column coordinate may be a non-integer rational number.

*lemma 5.7 within row index reduction trajectory cycle;*

There will be a cycle within a row, i.e. the Collatz Conjecture will be false, if any possible new start point lands on the boundary set by the prior row reduction trajectory.

$$\exists_{J>0}:\ (q(J)=(i_{0,}j_0))=(i_{0,}g^{-(i_0-1)}(J))$$

*lemma 5.8: monotonic divergence;*

Furthermore, it is sufficient to demonstrate divergence by showing that the start point will always be greater than the boundary set by the prior row reduction trajectory.

$$\forall_{J>0}:\ (q(J)=(i_{0,}j_0))>(i_{0,}g^{-(i_0-1)}(J))$$

----

These criteria can all be stated in terms of existential quantification:

*lemma 5.9;* $\neg\exists_{J>0}:\ (q(J)=(i_{0,}j_0))>(i_{0,}g^{-(i_0-1)}(J))$ monotonic convergence

*lemma 5.10;* $\exists_{J>0}:\ (q(J)=(i_{0,}j_0))=(i_{0,}g^{-(i_0-1)}(J))$ within boundary cycle

*lemma 5.11;* $\neg\exists_{J>0}:\ (q(J)=(i_{0,}j_0))<(i_{0,}g^{-(i_0-1)}(J))$ monotonic divergence

# 6 Start Point Behavior

In *lemma 5.3* for the definition of less than, on, and greater than relative to a boundary required comparison of the column coordinates along the row $i_0$ . As the row is given, we only need to calculate the column components of the comparison to understand the convergence behavior of the start point.

Of course, it is possible that none of our three criteria are true meaning that start points jump around the boundary. Such a result would tell us little about the truth of the Collatz Conjecture.

$g^{-1}(j)=(j-1)3^{-1}$ lemma 4.6

$g^{-2}(j)=((j-1)3^{-1}-1)3^{-1}$

$g^{-3}(j)=(((j-1)3^{-1}-1)3^{-1}-1)3^{-1}$

*lemma 6.1;* $g^{-m}(j)=(j-1)3^{-m}-\sum_{n=1}^{m-1}(3^{-n})$

Consider an example. note that $g^2(5)=49$ . Now lets apply 49 to $g^{-2}$ : $j+1=49$ . $j-1=48$ . The $(j-1)3^{-m}$ term is then $48\cdot3^{-2}=5\frac{1}{3}$ , a rational value. $m-1$ is just 1 so the summation term is $\frac{1}{3}$ , so we get the correct answer back of 5.

Note, 49 = $1210_3$ when this is shifted over two places we get $12.10_3$ Then when we subtract $0.1_3$ (a string of $m-1$ ones) we are left with $12_3$ , which is 5.

$$g^{-m}(j)=(j-1)3^{-m}-\sum_{n=1}^{m-1}(3^{-n})$$ *lemma 6.1*

$$g^{-m}(j)=3^{-m}[j-\sum_{n=1}^{m}(3^{m-n})]$$ pull out the $3^{-m}$

*lemma 6.2;* $$g^{-m}(j)=3^{-m}[j-\sum_{n=0}^{m-1}3^{n}]$$ fix the summation counter

*lemma 6.3;* $$g^{-m}(j)=3^{-m}[j-\frac{1}{2}(3^{m}-1)]$$ replace summation

Recall the restart function from *lemma 5.2*. I drop the subscripts here as there are no other *i, j* variables in this context, also these are just intermediate values used so the criteria equations will not become too long.

*def 6.1;* $i=\|tail(\gamma(J))\|$

*def 6.2;* $j=head(\gamma(J))$

and placing these into the convergence criteria of lemma 5.6, lemma 5.7, and lemma 5.8:

*lemma 6.3;* $\forall_J : j < D$ monotonic convergence

*lemma 6.4;* $\exists_J : j = D$ cycles within boundary

*lemma 6.5;* $\forall_J : j > D$ monotonic divergence

*lemma 6.6;* $$D=3^{-(i-1)}[J-\frac{1}{2}(3^{i-1}-1)]$$

All the criteria have the same right hand side, so we will call that the determinate, *D*.

Lets test a point to see if the column index reduces after a row index reduction, say

$J=40$ termination point for an iteration

$\rightarrow$

$\gamma(J)=121$

$\|tail(\gamma(J))\|$

$head(\gamma(J))=30$

$\rightarrow \quad 30<3^{0}[40-\sum_{n=0}^{-1}3^{n}]$ *lemma 6.3*

$\rightarrow \quad 30<120$ there is a column index reduction

It is true in general that when $tail(\gamma(J))$ is 1, lemma 6.3 reduces to $head(\gamma(J))<J$ , which will be true for all $\gamma(J)$ , because $\gamma(J)$ is always odd, and $head(\gamma(J))$ is gotten by subtracting a positive non-zero value from *J*, namely by subtracting $tail(\gamma(J))$ .

# 7 Change of Quantification Variables

We desire to switch the universal quantification from J in our convergence/cycles/diverges criteria of lemmas 6.3, 6.4, and 6.5 to *i* and *j*, where *i* is the tail length of $\gamma(J)$ , and j is the head of $\gamma(J)$ . By doing this we would remove $i=\|tail(\gamma(J))\|$ and $j=head(\gamma(J))$ from the criteria. Those are complicated non-linear functions so eliminating them would simplify analysis of the convergence criteria.

However we know from *lemma 1.2* that multiples of three are missing from the set $\{\forall_J \gamma(J)\}$ Multiples of three are most easily discussed in modulo 3. Head and tail are defined in modulo 2. As 2 and 3 are relatively prime 'digits' in base 2 and base 3 are uncorrelated. This is a sort of cryptographic problem. Hence, we would not expect to find an easy way to express how multiples of three missing in $\{\forall_J \gamma(J)\}$ would affect *head* and *tail*. Perhaps this is the crux of the difficulty in making conclusions about the Collatz conjecture.

I propose that we union $\{\forall_J \gamma(J)\}$ with the multiples of three so we can quantify on *i* and *j*. Lets call the set of odd multiples of three, ***T***, and then create $\{\forall_J \gamma(J)\} \cup \boldsymbol{T}$ will cause spurious results that are apparently randomly dispersed, but we have a simple way to test if a value is spurious, namely by assembling the head and the tail, and then applying the inverse $\gamma$ function to see if we get an integer.

For existential proofs we will have to test members from the set of results and show at least one of them can be generated from $\{\forall_J \gamma(J)\}$ rather than from ***T.*** If a universal criteria holds, then it holds for both the values that derived from $\{\forall_J \gamma(J)\}$ and those from ***T***. In which case it will not matter that we added the spurious values.

*def 7.1;* any odd natural number can be found in the set $\{\forall_J \gamma(J)\} \cup \boldsymbol{T}$

*lemma 7.2;* Any $n \in (\{\forall_J \gamma(J)\} \cup \boldsymbol{T})$ can be synthesized from two natural numbers $j$ and $i \geq 1$

We just use the construction technique given with the equation from def 1,
$c^{-1}(i,j) = j2^{i+1} + 2^i - 1$

*lemma 7.3;* from *def 7.1* and *lemma 7.2* we have derived that we may replace $\gamma(J)$ with a formulation in terms of *i* and *j*, though at the cost of introducing spurious values into our criteria.

$$\gamma(J) \rightarrow j2^{i+1} + 2^i - 1$$

---

Now if we find a function to calculate *J* given *j* and *i*, we can substitute this function in and remove *J* from the equation and universally quantify our criteria only on *j* and *i*. Here is a derivation of such a function:

*lemma 7.4;* $J = \gamma^{-1}(\gamma(J))$ identity

$J = \gamma^{-1}(j2^{i+1}+2^i-1)$ lemma 7.3, $\gamma(J) \rightarrow j2^{i+1}+2^i-1$

$\exists_k : J = ((j2^{i+1}+2^i-1)2^k-1)3^{-1}$ lemma 1.1, $\gamma_k^{-1}(n) = (n2^k-1)3^{-1}$

*lemma 7.5;* $\exists_k : J = (2^k(2^i(2j+1)-1)-1)3^{-1}$

For the gamma inverse we could not use the one operand form of lemma 3.7.3 as the inverse is taking us from the new start point back to row 1. That inverse explicitly doesn't apply to going to row 1.

Now is it really true that any value J can be calculated in by the expression on the right in lemma 7.5? Suppose we had a value, say, x-1, which could not be created with this expression, no matter what value we chose for *i* and *j*:

$$\forall_{j,i,k} : x \neq (2^k(2^i(2j+1)-1)-1)3^{-1}$$

$$3x+1 \neq 2^k(2^i(2j+1)-1)$$

$$2^{-k}(3x+1) \neq 2^i(2j+1)-1$$

$\gamma(x) \neq j2^{i+1}+2^i-1$ term on LHS $\gamma$ , expand RHS

The applying the Collatz function is always odd, and we already did the proof that any odd number can be constructed with the form on the right hand side in chapter 3. Therefore there can be no such x.

*lemma 7.5*: all possible J can be represented by picking *j, i,* and *k* in the formula, $J = (2^k(2^i(2h+1)-1)-1)3^{-1}$ Furthermore, due to the manner we derived J, J will have the correct mathematical relationship to our choice of i*, j,* and *k*.

We picked up *k* because the Collatz inverse is not unique without it. J is an integer, so we cannot use any possible *k*, rather *k* must be chosen so that $2^k(2^i(2j+1)-1)-1$ a multiple of 3. I.e.

$2^k(2^i(2j+1)-1)-1$ is a multiple of 3:

$$2^k(2^i(2j+1)-1)-1 = 0 \quad mod\ 3$$

$$\rightarrow \quad 2^k(2^i(2j+1)-1) = 1 \quad mod\ 3$$

Consider the following fact:

$$k \in \boldsymbol{E} \rightarrow 2^k \, mod\, 3 = 1$$

$$k \in \boldsymbol{O} \rightarrow 2^k \, mod\, 3 = 2$$

Consider the case *k* is even:

$$2^i(2j+1)-1=1 \ \ mod\ 3 \qquad\qquad k\in E$$

$$2^i(2j+1)=2 \ \ mod\ 3 \qquad\qquad k\in E$$

This two can be broken into two cases due to $2^i$

$$2j+1=2 \ \ mod\ 3 \qquad\qquad k\in E\wedge i\in E$$

$$\rightarrow \qquad 2j=1 \ \ mod\ 3 \qquad\qquad k\in E\wedge i\in E$$

This says that h must be a multiple of 3, but we said we were going to make *j* universally quantified. The contradiction here is that we started this derivation with *lemma 7.3*, $J=\gamma^{-1}(\gamma(J))$ which makes use of $\{\gamma(J)\}$ not $\{\gamma(J)\}\cup\boldsymbol{T}$ . As we continue to explore the other cases, we will find constraints that eliminate all members of T in this fashion, after this point we will not have to be concerned about having spurious results.

We will now derive a parametrization for each *i,j,k* that will facilitate universal quantification on the parameters. We will make the corresponding parameters *t,u,v.*

So for all cases we have

*lemma 7.6;* parametrization

| k | i | $2^k(2^i(2j+1)-1)=1 \ \ mod\ 3$ | parametrization |
|---|---|---|---|
| $k\in\boldsymbol{E}$ | $i\in\boldsymbol{E}$ | $2j=1 \ \ mod\ 3$ | $j=3t+2$ ; $i=2u$ ; $k=2v$ |
| $k\in\boldsymbol{E}$ | $i\in\boldsymbol{O}$ | $2(2j+1)=2 \ \ mod\ 3$ | |
| | | $1\cdot j+2=2 \ \ mod\ 3$ | |
| | | $j=0 \ \ mod\ 3$ | $j=3t$ ; $i=2u+1$ ; $k=2v$ |
| $k\in\boldsymbol{O}$ | $i\in\boldsymbol{E}$ | $2((2j+1)-1)=1 \ \ mod\ 3$ | |
| | | $2(2j)=1 \ \ mod\ 3$ | |
| | | $j=1 \ \ mod\ 3$ | $j=3t+1$ ; $i=2u$ ; $k=2v+1$ |
| $k\in\boldsymbol{O}$ | $i\in\boldsymbol{O}$ | $2(2(2j+1)-1)=1 \ \ mod\ 3$ | |
| | | $2j+2=1 \ \ mod\ 3$ | |
| | | $2j=2 \ \ mod\ 3$ | $j=3t+1$ ; $i=2u+1$ ; $k=2v+1$ |

So we can universally quantify on t, u and v, and not worry about spurious results, though we have to consider 4 cases.

*lemma 7.7;* Values of J in terms of *t*, *u*, and *v*.

| k | i | | |
|---|---|---|---|
| k | i | $3J=(2^k(2^i(2j+1)-1)-1)$ | *lemma 7.5* |
| $k\in\boldsymbol{E}$ | $i\in\boldsymbol{E}$ | $3J=2^{2v}(2^{2u}(6t+5)-1)-1$ | $j=3t+2$ ; $i=2u$ ; $k=2v$ |
| $k\in\boldsymbol{E}$ | $i\in\boldsymbol{O}$ | $3J=2^{2v}(2^{2u+1}(6t+1)-1)-1$ | $j=3t$ ; $i=2u+1$ ; $k=2v$ |
| $k\in\boldsymbol{O}$ | $i\in\boldsymbol{E}$ | $3J=2^{2v+1}(2^{2u}(6t+3)-1)-1$ | $j=3t+1$ ; $i=2u$ ; $k=2v+1$ |
| $k\in\boldsymbol{O}$ | $i\in\boldsymbol{O}$ | $3J=2^{2v+1}(2^{2u+1}(6t+3)-1)-1$ | $j=3t+1$ ; $i=2u+1$ ; $k=2v+1$ |

Now the criteria from lemmas 6.3, 6.4, and 6.5 become:

*lemma 7.8;* $\forall_{t,u,v}: j < D$

$$k\in\boldsymbol{E}\wedge i\in\boldsymbol{E}\rightarrow\forall_{t,u,v}:3t+2<D$$
$$k\in\boldsymbol{E}\wedge i\in\boldsymbol{O}\rightarrow\forall_{t,u,v}:3t<D$$
$$k\in\boldsymbol{O}\wedge i\in\boldsymbol{E}\rightarrow\forall_{t,u,v}:3t+1<D$$
$$k\in\boldsymbol{O}\wedge i\in\boldsymbol{O}\rightarrow\forall_{t,u,v}:3t+1<D$$

*lemma 7.9;* $\exists_{t,u,v}: j = D$

$$k\in\boldsymbol{E}\wedge i\in\boldsymbol{E}\rightarrow\exists_{t,u,v}:3t+2=D$$
$$k\in\boldsymbol{E}\wedge i\in\boldsymbol{O}\rightarrow\exists_{t,u,v}:3t=D$$
$$k\in\boldsymbol{O}\wedge i\in\boldsymbol{E}\rightarrow\exists_{t,u,v}:3t+1=D$$
$$k\in\boldsymbol{O}\wedge i\in\boldsymbol{O}\rightarrow\exists_{t,u,v}:3t+1=D$$

*lemma 7.10;* $\forall_{t,u,v}: j > D$

$$k\in\boldsymbol{E}\wedge i\in\boldsymbol{E}\rightarrow\forall_{t,u,v}:3t+2>D$$
$$k\in\boldsymbol{E}\wedge i\in\boldsymbol{O}\rightarrow\forall_{t,u,v}:3t>D$$
$$k\in\boldsymbol{O}\wedge i\in\boldsymbol{E}\rightarrow\forall_{t,u,v}:3t+1>D$$
$$k\in\boldsymbol{O}\wedge i\in\boldsymbol{O}\rightarrow\forall_{t,u,v}:3t+1>D$$

In every case the left hand side is of linear in t and contains no other variables.

*lemma 7.11;*
Now Applying lemma 7.7 to lemma 6.6, D, we have:

$k \in \boldsymbol{E}, i \in \boldsymbol{E}$ $\quad i=2u$ ; $k=2v$ $\quad 3J=2^{2v}(2^{2u}(6t+5)-1)-1$

$$\rightarrow D=3^{-(2u-1)}[(2^{2v}(2^{2u}(6t+5)-1)-1)3^{-1}-\frac{1}{2}(3^{2u-1}-1)]$$

$$\rightarrow 3^{2u-1}D=(2^{2v}(2^{2u}(6t+5)-1)-1)3^{-1}-\frac{1}{2}(3^{2u-1}-1) \quad \text{; mul by } 3^{2u-1}$$

$$\rightarrow 2\cdot 3^{2u}D=2^{2v+1}(2^{2u}(6t+5)-1)-3^{2u}+1 \quad \text{; mul by 6}$$

$$\rightarrow 2\cdot 3^{2u}D=(6t+5)2^{2(u+v)+1}-2^{2v+1}-3^{2u}+1 \quad \text{; expand}$$

$$\rightarrow (2D+1)3^{2u}=(6t+5)2^{2(u+v)+1}-2^{2v+1}+1 \quad \text{; add } 3^{2u} \text{ both sides}$$

$$\rightarrow (2D+1)3^{2u}=(6t+5)2^{2v+1}(2^{2u}-1)+1 \quad \text{; collect}$$

$$\rightarrow (2D+1)3^{2u}=(6t+5)2^{2u}2^{2v+1}-2^{2v+1}+1$$

$k \in \boldsymbol{E}, i \in \boldsymbol{O}$ $\quad i=2u+1$ ; $k=2v$ $\quad 3J=2^{2v}(2^{2u+1}(6t+1)-1)-1$

$$\rightarrow D=3^{-(2u)}[(2^{2v}(2^{2u+1}(6t+1)-1)-1)3^{-1}-\frac{1}{2}(3^{2u}-1)]$$

$$\rightarrow (2D+1)3^{2u+1}=(6t+5)2^{2v+1}(2^{2u+1}-1)+1$$

$k \in \boldsymbol{O}, i \in \boldsymbol{E}$ ; $i=2u$ ; $k=2v+1$ $\quad 3J=2^{2v+1}(2^{2u}(6t+3)-1)-1$

$$\rightarrow D=3^{-(2u-1)}[(2^{2v+1}(2^{2u}(6t+3)-1)-1)3^{-1}-\frac{1}{2}(3^{2u-1}-1)]$$

$$\rightarrow (2D+1)3^{2u}=(6t+5)2^{2v+2}(2^{2u}-1)+1$$

$k \in \boldsymbol{O}, i \in \boldsymbol{O}$ ; $i=2u+1$ ; $k=2v+1$ $\quad 3J=2^{2v+1}(2^{2u+1}(6t+3)-1)-1$

$$\rightarrow D=3^{-(2u-1)}[(2^{2v+1}(2^{2u+1}(6t+3)-1)-1)3^{-1}-\frac{1}{2}(3^{2u-1}-1)]$$

$$\rightarrow (2D+1)3^{2u+1}=(6t+5)2^{2v+2}(2^{2u+1}-1)+1$$

No multiplying by negative numbers was involved. Hence, when we unwrap each D we can then do the inverse operations on the left hand side of the criteria, in otherwords, the Ds above can be replaced with the left hand side directly.

*lemma 7.12;* $\forall_{t,u,v}: j < D$

$$k \in \boldsymbol{E} \wedge i \in \boldsymbol{E} \rightarrow \forall_{t,u,v}: D(6t+5)3^{2u} < (6t+5)2^{2v+1}(2^{2u}-1)+1$$
$$k \in \boldsymbol{E} \wedge i \in \boldsymbol{O} \rightarrow \forall_{t,u,v}: (6t+1)3^{2u+1} < (6t+5)2^{2v+1}(2^{2u+1}-1)+1$$
$$k \in \boldsymbol{O} \wedge i \in \boldsymbol{E} \rightarrow \forall_{t,u,v}: (6t+3)3^{2u} < (6t+5)2^{2v+2}(2^{2u}-1)+1$$
$$k \in \boldsymbol{O} \wedge i \in \boldsymbol{O} \rightarrow \forall_{t,u,v}: (6t+3)3^{2u+1} < (6t+5)2^{2v+2}(2^{2u+1}-1)+1$$

*lemma 7.13;* $\exists_{t,u,v}: j = D$

$$k \in \boldsymbol{E} \wedge i \in \boldsymbol{E} \rightarrow \exists_{t,u,v}: (6t+5)3^{2u} = (6t+5)2^{2v+1}(2^{2u}-1)+1$$
$$k \in \boldsymbol{E} \wedge i \in \boldsymbol{O} \rightarrow \exists_{t,u,v}: (6t+1)3^{2u+1} = (6t+5)2^{2v+1}(2^{2u+1}-1)+1$$
$$k \in \boldsymbol{O} \wedge i \in \boldsymbol{E} \rightarrow \exists_{t,u,v}: (6t+3)3^{2u} = (6t+5)2^{2v+2}(2^{2u}-1)+1$$
$$k \in \boldsymbol{O} \wedge i \in \boldsymbol{O} \rightarrow \exists_{t,u,v}: (6t+3)3^{2u+1} = (6t+5)2^{2v+2}(2^{2u+1}-1)+1$$

*lemma 7.14;* $\forall_{t,u,v}: j > D$

$$k \in \boldsymbol{E} \wedge i \in \boldsymbol{E} \rightarrow \forall_{t,u,v}: (6t+5)3^{2u} > (6t+5)2^{2v+1}(2^{2u}-1)+1$$
$$k \in \boldsymbol{E} \wedge i \in \boldsymbol{O} \rightarrow \forall_{t,u,v}: (6t+1)3^{2u+1} > (6t+5)2^{2v+1}(2^{2u+1}-1)+1$$
$$k \in \boldsymbol{O} \wedge i \in \boldsymbol{E} \rightarrow \forall_{t,u,v}: (6t+3)3^{2u} > (6t+5)2^{2v+1}(2^{2u}-1)+1$$
$$k \in \boldsymbol{O} \wedge i \in \boldsymbol{O} \rightarrow \forall_{t,u,v}: (6t+3)3^{2u+1} > (6t+5)2^{2v+2}(2^{2u+1}-1)+1$$

Lets look at monotonic column convergence first. As *v* is a universally quantified variable, it appears only on the right side of the greater than comparison, and the right hand side becomes monotonically larger when v becomes larger, so we set the strongest right hand side constraint by setting *v* to zero:

*lemma 7.15;* $\forall_{t,u,v}: j < D$

$$k \in \boldsymbol{E} \wedge i \in \boldsymbol{E} \rightarrow \forall_{t,u,v}: 2^{\frac{\ln(3)}{\ln(2)}2u} < 2^{2u+1}-1$$
$$k \in \boldsymbol{E} \wedge i \in \boldsymbol{O} \rightarrow \forall_{t,u,v}: (6t+1)2^{\frac{\ln(3)}{\ln(2)}(2u+1)} < (6t+5)(2^{2u+2}-2)+1$$
$$k \in \boldsymbol{O} \wedge i \in \boldsymbol{E} \rightarrow \forall_{t,u,v}: (6t+3)2^{\frac{\ln(3)}{\ln(2)}2u} < (6t+5)(2^{2u+2}-4)+1$$
$$k \in \boldsymbol{O} \wedge i \in \boldsymbol{O} \rightarrow \forall_{t,u,v}: (6t+3)2^{\frac{\ln(3)}{\ln(2)}(2u+1)} < (6t+5)(2^{2u+3}-4)+1$$

It is interesting that *t* has little or no role.

In all cases the power of u is larger on the left hand side than on the right hand side, so the constraint can not hold universally. It follows that:

*Theorem 7.1;* No Monotonic Column convergence algorithm based on the Cadogan function exists.

Lets take a look at divergence.

*lemma 7.14;* $\forall_{t,u,v}: j > D$

$$k\in\boldsymbol{E}\wedge i\in\boldsymbol{E}\rightarrow\forall_{t,u,v}:(6t+5)3^{2u}>(6t+5)2^{2v+1}(2^{2u}-1)+1$$
$$k\in\boldsymbol{E}\wedge i\in\boldsymbol{O}\rightarrow\forall_{t,u,v}:(6t+1)3^{2u+1}>(6t+5)2^{2v+1}(2^{2u+1}-1)+1$$
$$k\in\boldsymbol{O}\wedge i\in\boldsymbol{E}\rightarrow\forall_{t,u,v}:(6t+1)3^{2u+1}>(6t+5)2^{2v+1}(2^{2u+1}-1)+1$$
$$k\in\boldsymbol{O}\wedge i\in\boldsymbol{O}\rightarrow\forall_{t,u,v}:(6t+3)3^{2u+1}>(6t+5)2^{2v+2}(2^{2u+1}-1)+1$$

As *v* is a free variable on the RHS it can be set to an unbounded number of values to create exceptions to the constraints, no matter how *t* and *u* are set.

Now lets take a look at within trajectory cycles. This constraint is fundamentally different because it only relates numbers found within a single row index reduction trajectory. The other two constraints spoke of relationships between trajectories.

*lemma 7.13;* $\exists_{t,u,v}: j = D$

$$k\in\boldsymbol{E}\wedge i\in\boldsymbol{E}\rightarrow\exists_{t,u,v}:(6t+5)3^{2u}=(6t+5)2^{2v+1}(2^{2u}-1)+1$$
$$k\in\boldsymbol{E}\wedge i\in\boldsymbol{O}\rightarrow\exists_{t,u,v}:(6t+1)3^{2u+1}=(6t+5)2^{2v+1}(2^{2u+1}-1)+1$$
$$k\in\boldsymbol{O}\wedge i\in\boldsymbol{E}\rightarrow\exists_{t,u,v}:(6t+1)3^{2u+1}=(6t+5)2^{2v+1}(2^{2u+1}-1)+1$$
$$k\in\boldsymbol{O}\wedge i\in\boldsymbol{O}\rightarrow\exists_{t,u,v}:(6t+3)3^{2u+1}=(6t+5)2^{2v+2}(2^{2u+1}-1)+1$$

$$(6t+5)2^{\frac{ln3}{ln2}2u}=(6t+5)2^{2v+1}(2^{2u}-1)+1 \qquad \text{case 0}$$
$$3^{2u}-2^{2(u+v)+1}-2^{2v+1}=1/(6t+5)$$

As all the terms on the left are integers, and the term on the right is a fraction, there are no solutions to this equation. Lets try case 1:

$$(6t+1)3^{2u+1}=(6t+5)2^{2v+1}(2^{2u+1}-1)+1 \qquad \text{case EO}$$
$$(6t+1)3^{2u+1}=(6t+5)2^{2(u+v+1)}-(6t+5)2^{2v+1}+1$$
$$3^{2u+1}+2^{2v+1}=2^{2(u+v+1)}+\frac{4\cdot3^{2u+1}+1}{6t+5}$$
$$3^{2u+1}+2^{2v+1}-2^{2(u+v+1)}=\frac{4\cdot3^{2u+1}+1}{6t+5}$$
$$n(6t+5)=4\cdot3^{2u+1}+1 \qquad \text{looking at fraction, it must be an integer}$$
$$3^{2u+1}+2^{2v+1}>2^{2(u+v+1)} \qquad \text{term on left must remain positive}$$

What a nice equation, a mixed integer exponential problem in two bases along with factoring.

The other cases have a similar form based on small constant transformations of *u, v, and t.*

## Conclusion

The head of an arbitrary number is another arbitrary number; hence it is difficult for me to see how stripping off the tail helps us gain understanding of the convergence of the Collatz function. However, I do find interesting that strands of numbers can be easily identified in the sequence where *k=1*, and algebraic functions may be used. Analysis of these strands lead to a closed form description for possible cycles in them. However, evaluating the form requires large integer factorization and mixed integer calculation.

## The Reference

## Appendix

Here is a Mathematica program for evaluating the following form,

$$3^a - 2^b(2^a - 1) = \frac{4 \cdot 3^a + 1}{c}$$

There is a solution at {a = 1, b=1, c=13} but these are not the correct form for the parameters, e.g. no 6 t + 5 = 13.